\documentclass[12pt]{iopart}
\usepackage{epsfig}
\usepackage{bm}
\usepackage[matrix,frame,arrow]{xy}
\usepackage{color}
\usepackage{amssymb}
\usepackage[mathscr]{eucal}
\usepackage{tensor}
\usepackage{bbm}
\usepackage{amsfonts}
\usepackage{subfigure}
\usepackage{graphicx,cite}
\usepackage{psfrag}
\usepackage{pstool}
\usepackage{iopams}

\newcommand{\bea}{\begin{eqnarray}}
\newcommand{\eea}{\end{eqnarray}}
\newcommand{\be}{\begin{equation}}
\newcommand{\ee}{\end{equation}}
\newcommand{\ls}{\left[}
\newcommand{\rs}{\right]}
\newcommand{\lr}{\left(}
\newcommand{\rr}{\right)}

\newcommand{\bsm}{\begin{bmatrix}}
\newcommand{\esm}{\end{bmatrix}}
\newcommand{\bin}[2]{{{#1} \choose {#2}}}
\begin{document}

\title[]{Scattering of coherent pulses on a two-level system -- single-photon generation}

\author{J. Lindkvist$^1$, G. Johansson$^2$}

\address{Department of Microtechnology and Nanoscience, Chalmers University of Technology, G\"oteborg, Sweden}
\ead{$^1$joell@chalmers.se}
\ead{$^2$Goran.L.Johansson@chalmers.se}
\begin{abstract}
In this work, we consider a two-level system (TLS) coupled to a one-dimensional continuum of bosonic modes in a transmission line. Using the master equation approach, a method for determining the photon number distribution of the scattered field is outlined. Specifically, results for the reflected field when driving the TLS with a coherent pulse are given. While the one-photon probability is enhanced compared to the incident coherent field, the system is still not a good deterministic single-photon source. Extending the system to contain two separate transmission lines, however, output fields with one-photon probabilities close to unity can be reached.

\end{abstract}

\section{Introduction}
In recent years, experimental advances in the field of circuit quantum electrodynamics (cQED) \cite{clarke08,schoelkopf08,Devoret13} have opened up possibilities for studying light-matter interactions in a wide range of coupling regimes, including the ultrastrong regime \cite{FornDiaz10,niemczyk10}. In cQED, artificial atoms based on Josephson junctions interact with electromagnetic fields in superconducting circuits, typically in the microwave regime. The advances in this research area have largely been driven by potential applications in quantum computing, where the artificial atoms play the role of qubits and the photons are used for qubit manipulation. While most of the experiments focus on fields confined to a cavity, there is now a growing interest in systems where the artificial atom interacts with a continuum of propagating photonic modes in an open transmission line. These systems have potential applications in optical quantum information \cite{knill01}, where the roles of the photons and the artificial atoms are reversed. The information is carried by the photons, while the atoms are used to engineer environments for manipulating the photons and mediating effective multi-photon interactions. 

Systems with (artificial) atoms coupled to a one-dimensional continuum of photonic modes have been theoretically analyzed in \cite{shen05, shen05b, PhysRevA.82.063816, PhysRevA.85.043832, PhysRevA.85.023817, peropadre13, dibyendu10, dibyendu11} and experimental impementations in cQED include \cite{Astafiev10, AstafievPRL2010, AbdumalikovPRL2010, AbdumalikovPRL2011,hoi11, hoi12, Hoi122, Hoi123,vanLoo12,Eichler12}. In order to use propagating photons for quantum information, several tools for manipulation, like single-photon sources and detectors, are needed. Today, the standard method for detection and state reconstruction of propagating microwave fields builds on linear amplifiers \cite{menzel10,blais10}. Single-photon detection has been theoretically analyzed in e. g. \cite{romero09,peropadre11,fan13,sankar13}. Moreover, single-photon sources utilizing transmission line resonators have been experimentally realized \cite{houck07,pechal13,lang13,yin13}.

In this paper, we examine the possibilities of constructing a non-cavity based single-photon source. Resonant scattering of coherent states on a two-level system (TLS) has been shown theoretically \cite{PhysRevA.82.063816,PhysRevA.85.023817,peropadre13,lukin07} and experimentally \cite{hoi12} to result in a photon number redistribution, with the single-photon probability being enhanced (suppressed) in the reflected (transmitted) field. The idea is to exploit this fact to, by choosing an appropriate input pulse, use the TLS as an on-demand single-photon source. Previous treatments of coherent state scattering on a TLS have been restricted to low-power input pulses \cite{PhysRevA.82.063816} or constant input fields of arbitrary intensity \cite{peropadre13,PhysRevA.85.023817}. For our purposes, we extend the analysis to pulses of arbitrary intensity, building on the results for continuous coherent driving in reference \cite{peropadre13}. While our main motivation for this work is cQED implementations, the analysis is more general and applies to any system described by a TLS coupled to a continuum of photonic modes in one dimension. A different example is the surface plasmon systems \cite{lukin07}.

The paper is organized as follows. In section \ref{sec:master}, we introduce our system and write down the master equation for the TLS and expressions for the output field operators. In section \ref{sec:probabilities}, we show how to determine the photon number distribution, given the master equation and the field operators. In section \ref{sec:distribution}, results for the reflection of coherent square pulses are given. Finally, in section \ref{sec:twolines}, we extend the setup to include two transmission lines and determine the photon number distribution in the same way.

\section{System and master equation}
\label{sec:master}
\begin{figure}[ht]
\centering
\includegraphics[width=0.5 \columnwidth]{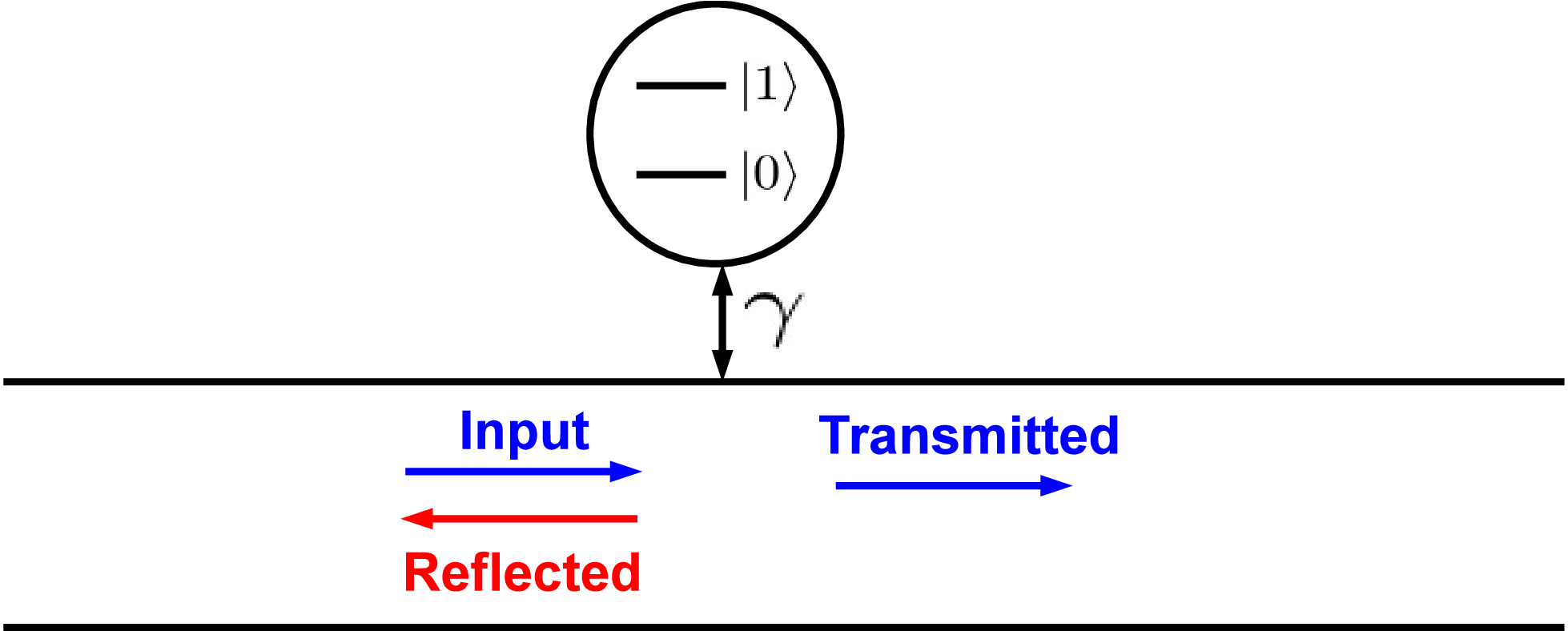}
\caption{Sketch of the setup considered. A two-level system is coupled to a continuum of left- and right-propagating bosonic modes with coupling strength $\gamma$. }
\label{fig:sketch}
\end{figure}

We consider a two-level system (TLS) locally coupled to an infinite transmission line (TL), i. e. a 1D open space supporting a continuum of left- and right-propagating photonic modes.  A schematic sketch of the setup is shown in figure \ref{fig:sketch}. The coupling $\gamma$ is assumed to be strong in the sense that the damping of the TLS is dominated by the relaxation to the TL. One particular example of such a system, inspiring us to this work, is a transmon qubit in the two-level approximation coupled to a superconducting coplanar waveguide. This system was theoretically analyzed in \cite{peropadre13} and has been explored in recent experiments with propagating microwave photons \cite{hoi12}.
The configuration we consider can be described in terms of a quantum Langevin equation (QLE) for the TLS coupled to a bath of harmonic oscillators, togheter with input-output equations relating the incoming and outgoing states at the interaction point \cite{gardinerzoller}. The TLS operators are described in terms of Pauli matrices. 
We normalize all our photon fields $V(t)$ so that the first-order correlation function,
\be
G^{(1)}(t)=\langle V^-(t)V^{+}(t)\rangle,
\ee
is equal to the number of photons per relaxation time $1/\gamma$. Here $V^+(t)$ and $V^-(t)$ denote the positive and negative frequency parts, respectively. In a superconducting coplanar waveguide, $V(t)$ is proportional to the voltage between the center conductor and ground.

We consider the case where we have a coherent driving signal $V_{in}(t)$ from the left, close to resonance with the TLS, and vaccum as input from the right. With $\sigma^x(t)=\sigma^+(t)+\sigma^-(t)$ being the TLS operator coupling to the TL, the input-output equations result in the following expressions for the transmitted (right) and reflected (left) outgoing fields,
\bea
V_T^{\pm}(t)=V_{in}^{\pm}(t)+\frac{1}{\sqrt{2}}\sigma^{\mp}(t)\\
V_R^{\pm}(t)=\frac{1}{\sqrt{2}}\sigma^{\mp}(t).\label{reflected}
\eea
We take our coherent input signal to be of the form
\be
\langle V_{in}(t)\rangle=A(t)\sin{\omega_dt},
\ee
where $A(t)$ is the pulse amplitude envelope function, varying on a timescale much slower than $1/\omega_d$.

From the quantum Langevin equation, we can derive a master equation for the reduced density matrix, $\rho(t)$, of the TLS.  We assume that the full density matrix initially can be written as a direct product, that the coupling $\gamma$ is much smaller than the energy of the TLS and that the correlation times of the TL variables are short compared to other timescales. Neglecting thermal excitations, in a frame rotating with the driving frequency $\omega_d$ and after employing the rotating-wave approximation, the master equation reads
\be
\dot{\rho}(t)=i\frac{\Delta}{2}\ls\sigma^z,\rho(t)\rs+˜\gamma\mathcal{D}(\sigma^-)\rho(t)-i\gamma A(t)\sqrt{\frac{\omega_d}{8\omega_{10}}}\ls\sigma^x,\rho(t)\rs,
\label{eq:master}
\ee
where $\omega_{10}$ is the transition frequency of the TLS and $\Delta\equiv\omega_{10}-\omega_d$ is the detuning. The Lindblad operator is defined by $\mathcal{D}(c)\rho=c\rho c^{\dagger}-\frac{1}{2}\lr c^{\dagger}c\rho+\rho c^{\dagger}c\rr$. With a resonant drive ($\Delta=0$) and time expressed in units of the relaxation time $1/\gamma$, the master equation takes the simple form
\be
\dot{\rho}(t)=\mathcal{D}(\sigma^-)\rho(t)-i\sqrt{\frac{N_{in}(t)}{2}}\ls\sigma^x,\rho(t)\rs,
\label{MEsimplified}
\ee
where $N_{in}(t)$ is the number of incoming photons per relaxation time. Symbolically, we can write the master equation as
\be
\dot{\rho}(t)=L(t)\rho(t),
\label{MEformal}
\ee
where $L(t)$ denotes the Liouvillian super-operator.

\section{Photon number distribution}
\label{sec:probabilities}
 In this section, we outline how to fully determine the photon number distribution from the master equation and the field operator. This procedure applies to any 1D photon field, but will in section \ref{sec:distribution} be applied to the field reflected from a TLS driven by a coherent pulse.
Our task is to compute all the probabilities $P_n$ for the field to contain $n$ photons, up to some cutoff $k$ where it becomes negligible. 
The average number of photons in the field is given by $N_1=\sum_nn P_n$ and the average number of photon pairs by $N_2=\sum_n\bin{n}{2}    P_n$. More generally, the number of photon $m$-tiples can be expressed as
\be
N_m=\sum_{n=m}^{k}\bin{{n}}{{m}} P_n.
\label{eq:mtiples}
\ee
With knowledge of all $m$-tiple numbers $N_m$, equation (\ref{eq:mtiples}) can be inverted in order to obtain all the probilities $P_n$.

To determine $N_m$ we need to compute the $m$:th order time-ordered correlation function, defined by
\be
G^{(m)}(t_1,t_2,...,t_m)=\langle V^-(t_1)...V^-(t_m)V^+(t_m)...V^+(t_1)\rangle.
\ee
With our field normalized as in the previous section, $G^{(m)}(t_1,t_2,...,t_m)$ is equal to the photon coincidence rate at times $t_1,...,t_m$. Thus, in a time interval $[t_i,t_f]$, the number of photon $m$-tiples is given by
\be
N_m=\int_{t_i}^{t_f}dt_1\int_{t_1}^{t_f}dt_2...\int_{t_{m-1}}^{t_f}dt_m{G}^{(m)}(t_1,...,t_m).
\label{eq:Mn}
\ee
In order to compute the correlation functions we need the density matrix $\rho (t)$ and the two-time propagator super-operator $P(t,t_0)$, defined by $\rho(t)=P(t,t_0)\rho (t_0)$. It follows from its definition and equation (\ref{MEformal}) that the propagator satisfies the equation
\be
\dot{P}(t,t_0)=L(t)P(t,t_0),
\label{eq:prop}
\ee
with the initial condition $P(t_0,t_0)=1$.

In terms of the density matrix and the propagator, the $m$:th order correlation function is given by \cite{gardinerzoller}
\be
G^{(m)}(t_1,t_2,...,t_m)=\tr{\ls\hat{n}P(t_m,t_{m-1})...\hat{n}P(t_2,t_1)\hat{n}\rho (t_1)\rs},
\label{eq:Gm}
\ee
where $\hat{n}$ is defined in terms of the field Schr\"odinger operator $V$ as $\hat{n}\rho=V^+\rho V^-$ and the propagator acts on everything to the right. For any Liouvillian $L(t)$ we can solve equations (\ref{MEformal}) and (\ref{eq:prop}) for the density matrix and the propagator. Given an expression for the field operator $V$, equation (\ref{eq:Gm}) can then be used to compute all the correlation functions $G^{(m)}(t_1,t_2,...,t_m)$, allowing us to determine the photon number probability distribution via equations (\ref{eq:mtiples}) and (\ref{eq:Mn}). In particular, we will carry out this procedure for the field reflected from a TLS driven by a time-dependent coherent signal. In this case, $V$ and $L(t)$ are given by equations  (\ref{reflected}) and (\ref{MEsimplified}).

\section{Distribution for the reflected field}
\label{sec:distribution}
As mentioned in the introduction, earlier calculations and experiments have shown that the one-photon probability is enhanced in the reflected field when driving the TLS with a constant coherent signal.
Since our main motivation for this work is to investigate the possibility of an on-demand single-photon source, we compute the photon number distribution for the reflected field when driving the TLS with a coherent pulse. 

For a given pulse shape, there are two free parameters in the problem; the temporal width of the pulse $T$ (in units of the relaxation time) and the total mean number of photons in the pulse $N$. In the following we restrict the analysis to square pulses. These are easy to generate in microwave superconducting circuits and therefore convenient to use as input. Moreover, the master equation (\ref{MEsimplified}) can be solved analytically for square pulses.

\begin{figure}[!ht]
\centering
 \includegraphics[width=0.5 \columnwidth]{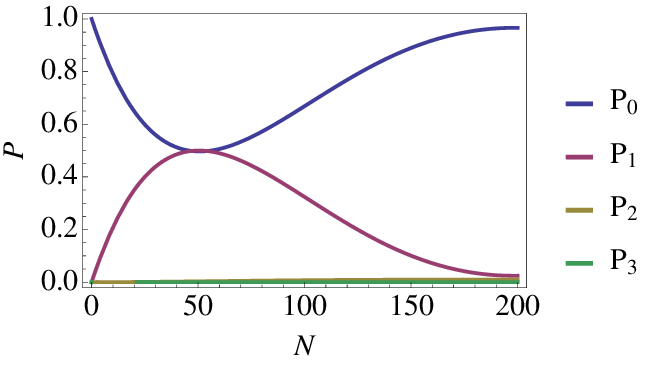}
\caption{The lowest photon number probabilities for the reflection of a coherent square pulse with width $T=0.1$ as a function of the mean number of incoming photons $N$.} 
\label{fig:shortpulse}
\end{figure}

Solving the master equation and carrying out the procedure outlined in the previous section, we end up with analytic, but complicated, expressions for the lowest photon number probabilities. 
\begin{figure}[ht]
\centering
\subfigure{
	\includegraphics[width=0.4\columnwidth]{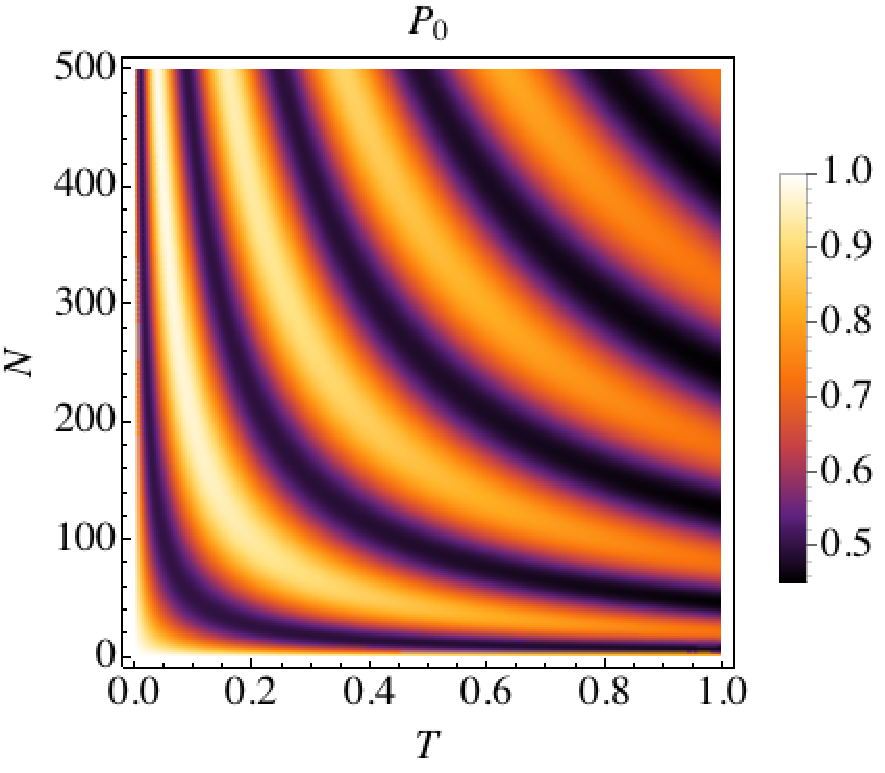}
}
\subfigure{
	\includegraphics[width=0.4\columnwidth]{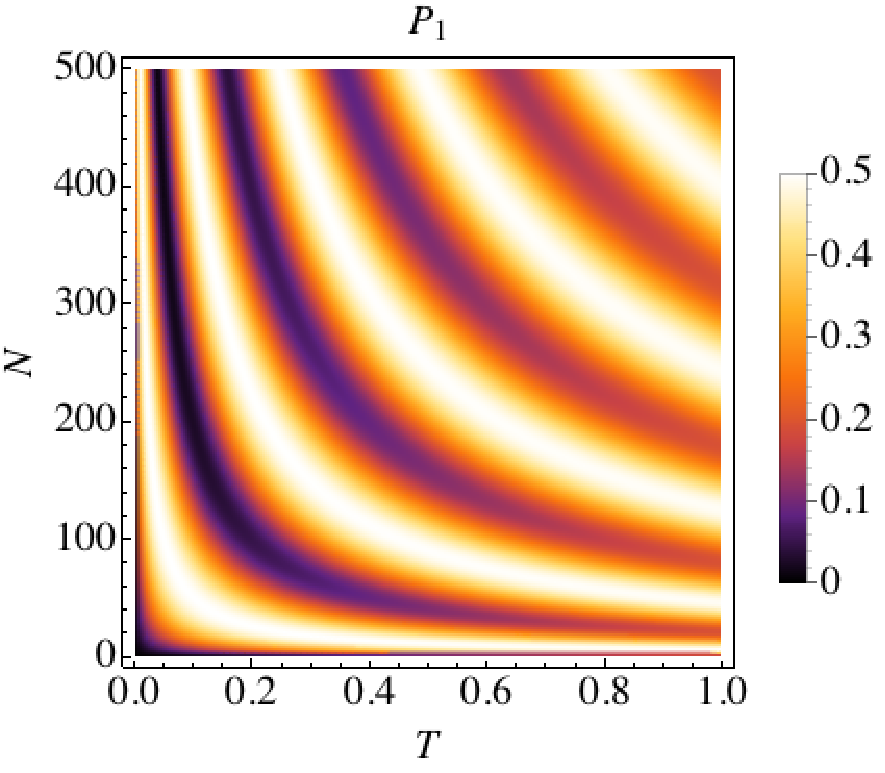}
}\\
\subfigure{
	\includegraphics[width=0.4\columnwidth]{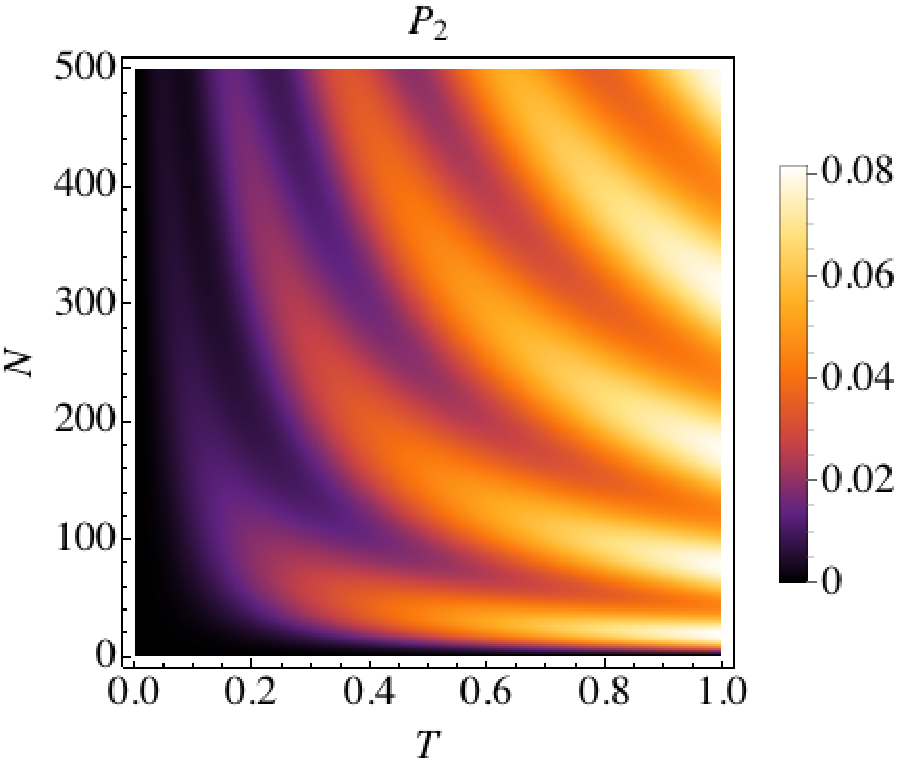}
}
\subfigure{
	\includegraphics[width=0.4\columnwidth]{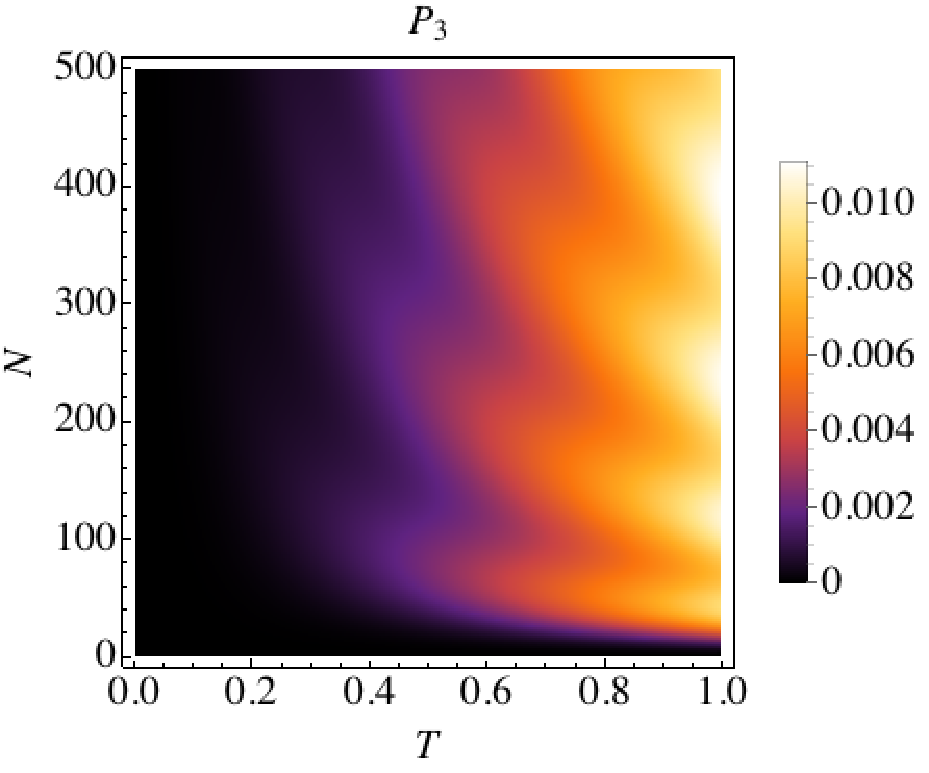}
}
\caption[]{Photon number probabilities for the field reflected from a TLS as a function of pulse width $T$ and number of incoming photons $N$.}
\label{fig:colorplots}
\end{figure}
In figure \ref{fig:colorplots} we plot $P_0$, $P_1$, $P_2$ and $P_3$ as a function of $T$ and $N$. For a given pulse width, the one-photon probability periodically reaches a maximal value of $P_1\approx0.5$. When the pulse width starts to become comparable to the relaxation time, however, the probability to scatter more than one photon becomes non-negligible. In order to maximize the one-photon probability while keeping the higher photon number probabilities low, we should look at short pulses. Figure \ref{fig:shortpulse} shows the probabilities for $T=0.1$, a regime where the probability to scatter more than one photon is negligible. For increased input powers, the probability to scatter a photon increases until is reaches a maximal value of $0.5$. For larger input powers, the Rabi splitting of the transition leads again to a lower scattering probability. The optimal point essentially corresponds to exciting the TLS with a $\pi$-pulse and letting it relax with probability $0.5$ in each direction. For the purposes of an on-demand single-photon source, this is the best we can do with a TLS coupled to a single transmission line. In section \ref{sec:twolines}, we consider an extended setup, where the TLS is coupled to an additional transmission line.

\section{Two transmission lines}
\label{sec:twolines}
\begin{figure}[ht]
\centering
\includegraphics[width=0.5 \columnwidth]{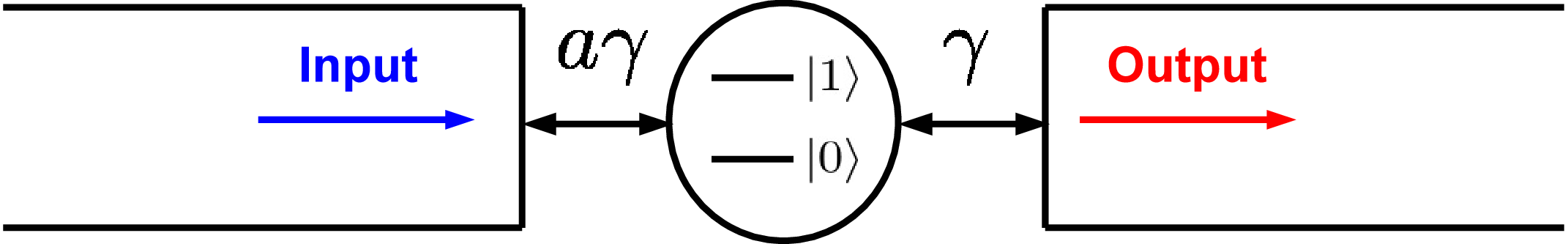}
\caption{Sketch of the setup with two semi-infinite transmission lines. The ratio of the couplings to the two lines is $a\le1$. The weakly coupled line is used to excite the TLS and a photon leaks out into the strongly coupled line with high probability.}
\label{fig:sketch2}
\end{figure}

In this section, we consider an extension of the previous setup, where the TLS is coupled to two semi-infinite transmission lines (see figure \ref{fig:sketch2}). As an example, the circuit analysis in \cite{peropadre13} of a transmon coupled to a coplanar waveguide can be generalized to accommodate this case as well. In the extended setup, there is an additional free parameter; the ratio of the couplings to the two transmission lines. The main idea is to couple the TLS much more strongly to one of the lines. After exciting it through the weakly coupled line, a photon will be emitted to the other one with high probability.

\begin{figure}[ht]
\centering
\subfigure{
	\includegraphics[width=0.4\columnwidth]{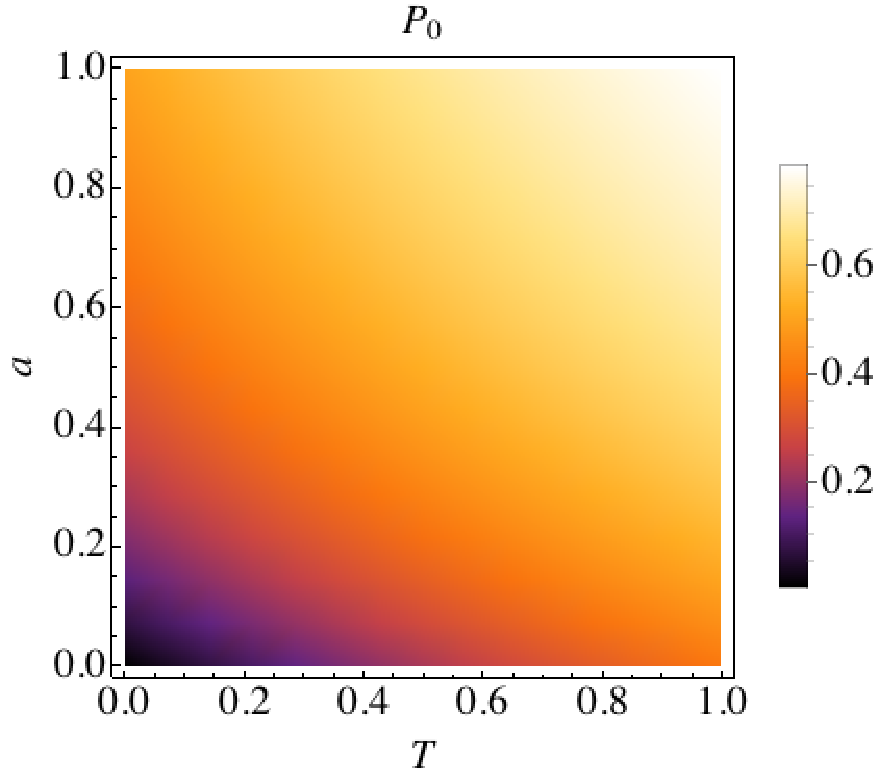}
}
\subfigure{
	\includegraphics[width=0.4\columnwidth]{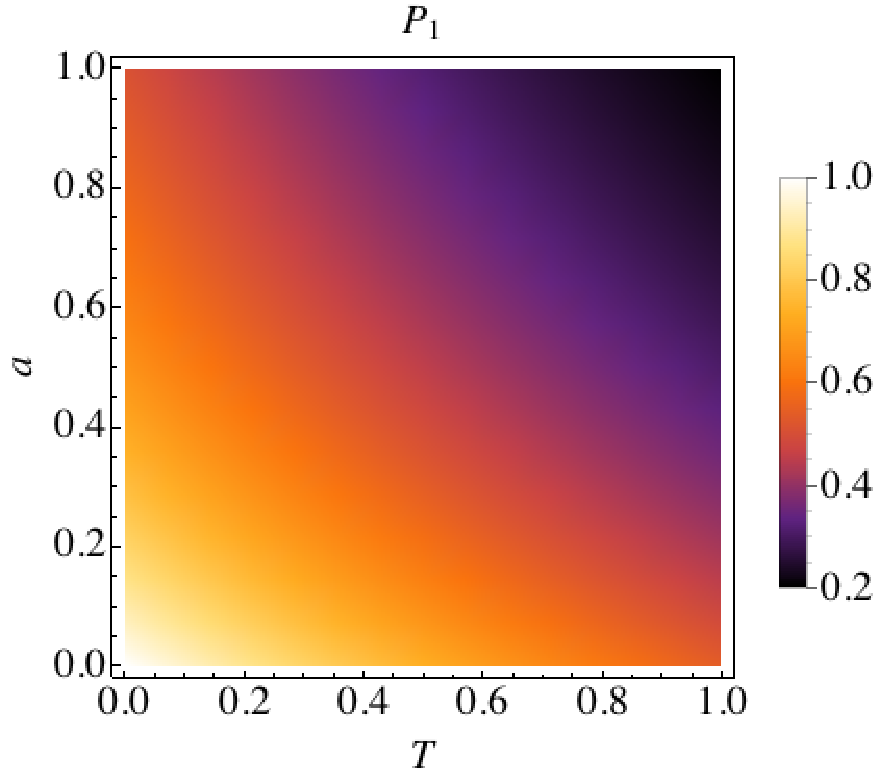}
}\\
\subfigure{
	\includegraphics[width=0.4\columnwidth]{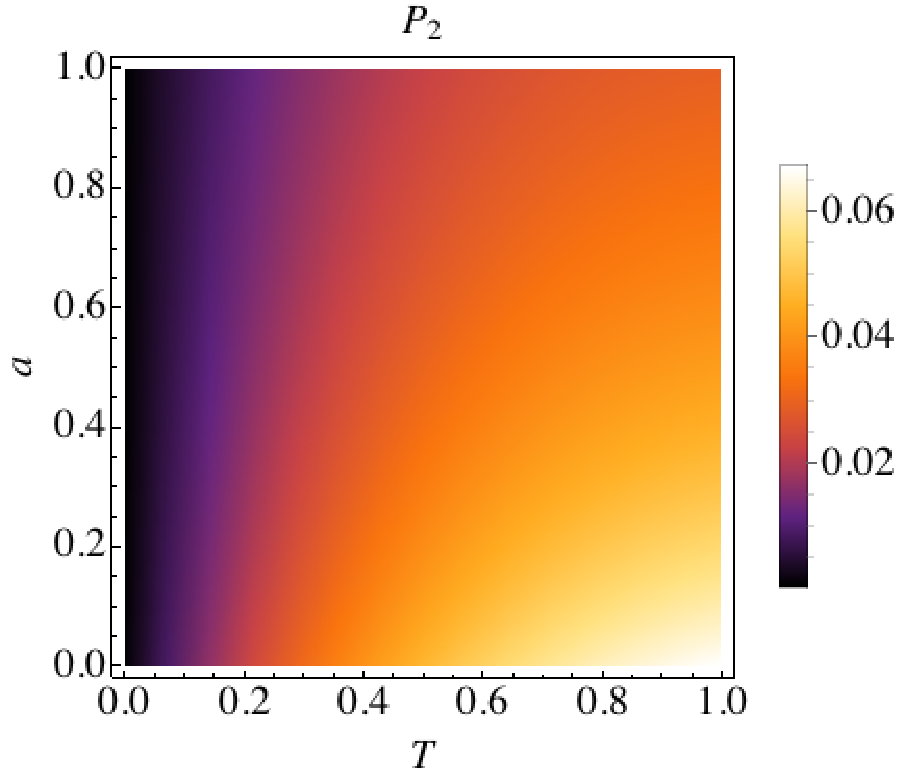}
}
\subfigure{
	\includegraphics[width=0.4\columnwidth]{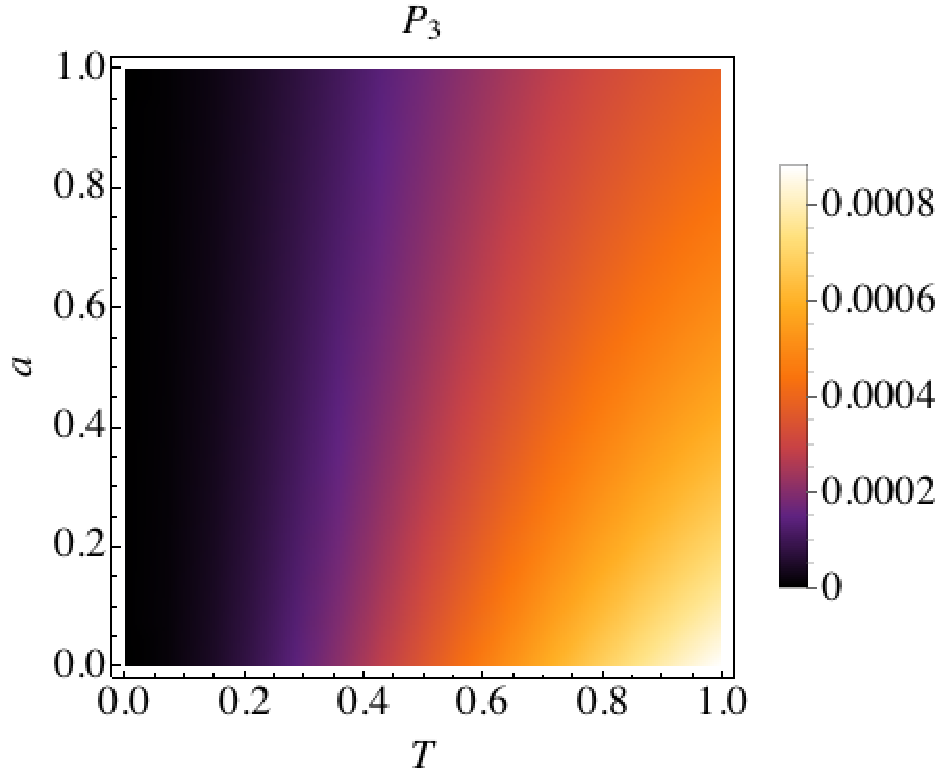}
}
\caption[]{Photon number probabilities for the output field in the strongly coupled transmission line as a function of the pulse width $T$ and the coupling ratio $a$, for the value of $N$ where $P_1$ is at its maximum. In the limit $T\rightarrow 0$, the probabilities are given by $P_1=1/(1+a)$, $P_0=1-P_1$ and $P_2=P_3=0$.}
\label{fig:colorplots2}
\end{figure}

The configuration can be described in terms of a quantum Langevin equation for a TLS coupled to two separate baths. With a coherent time-dependent drive in the weakly coupled line and vacuum as input in the strongly coupled line, the master equation takes the following form
\be
\dot{\rho}=(1+a)\mathcal{D}(\sigma^-)\rho-i\sqrt{aN_{in}(t)}\ls\sigma^x,\rho\rs,
\label{MEsimplified2}
\ee
where $a\le 1$ is the coupling ratio. Time is expressed in units of the relaxation time into the strongly coupled line. The positive and negative frequency parts of the output field in this line are $V_S^{\pm}(t)=\sigma_{\mp}(t)$.
\begin{figure}[ht]
\centering
\subfigure{
	\includegraphics[width=0.47\columnwidth]{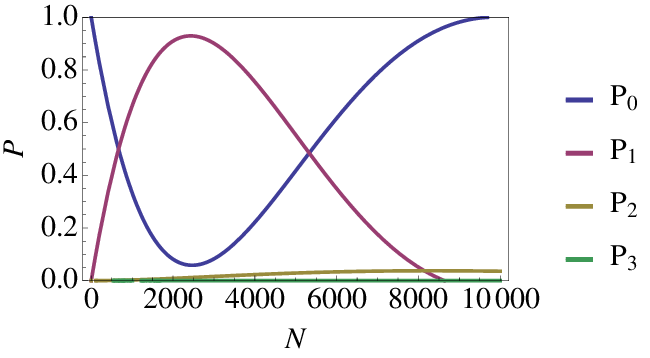}
}
\subfigure{
	\includegraphics[width=0.45\columnwidth]{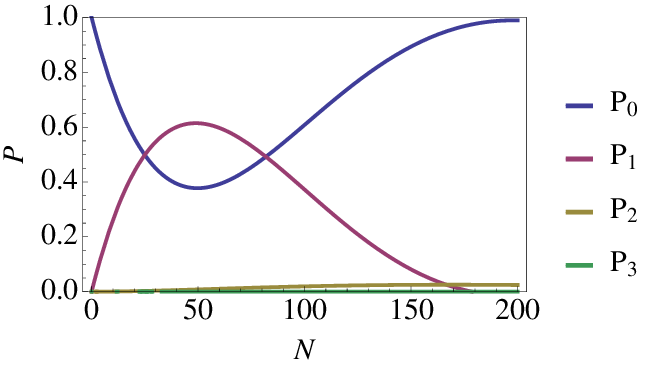}
}
\caption[]{Photon number probabilities for the output field in the strongly coupled line for $T=0.1$. The coupling ratios are $a=0.01$ (left) and $a=0.5$ (right).}
\label{fig:twolineplots}
\end{figure}
In figure \ref{fig:twolineplots} we plot the photon number probabilities for $T=0.1$, with two different coupling ratios. By choosing smaller values of $a$, we can now reach higher one-photon probabilities in the output field. Figure \ref{fig:colorplots2} shows the maximal values of $P_1$ as a function of $a$ and $T$, together with the corresponding values of $P_0$, $P_2$ and $P_3$. For small $a$ and $T$ we see that $P_1$-values close to unity can be reached. 
How small these parameters can be in practice depends on the specific physical realization. In our example of a transmon coupled to a waveguide, the limited anharmonicity of the transmon sets a limit on the pulse width. The results, however, depend on $T$, which is the pulse width in units of relaxation time (inverse coupling). One can thus improve the results by decreasing the coupling to the output line, but this results in a long output photon, which may not always be desirable in applications. Essentially, there is a trade-off between how short we want the output photon to be and how deterministic the single-photon source is.  For a typical anharmonicity of $500$ MHz we could, as an example, obtain $P_1=97\%$ and $P_2=0.6\%$ using a coupling of $4$ MHz.

Finally, in the regime where a deterministic single-photon source is realistic, we can also use a modification of the setup to create a source of single-photon path entangled states. Replacing the semi-infinite strongly coupled output line with an infinite transmission line, the photon will leak out symmetrically in both directions.
Denoting the zero- and one-photon states in the left (right) direction by $|0\rangle_L$ and $|1\rangle_L$ ($|0\rangle_R$ and $|1\rangle_R$), this would produce the state
\be
|\psi\rangle=\frac{1}{\sqrt{2}}\lr |0\rangle_L|1\rangle_R+|1\rangle_L|0\rangle_R\rr.
\label{entangled}
\ee
This is an entangled state that could be used to carry out a Bell test for microwave photons \cite{morin13}. In principle, the setup with a single transmission line could also be used to generate the state in equation (\ref{entangled}). However, in this case, one of the output ports would contain also the large transmitted part of the input pulse. Although separated in time from the entangled state, this limits the usefulness of the setup.

\section{Summary and conclusions}
To summarize, we have analyzed the interaction of a propagating photon field in a one-dimensional waveguide with a two-level system. In particular, we have determined the photon number distribution for the reflected field when driving the TLS with a coherent pulse. This setup is not a suitable on-demand single-photon source, since one-photon probabilities higher than 0.5 cannot be obtained. We have also considered a generalized setup with two transmission lines, allowing us to achieve one-photon probabilities close to unity in certain cases. Since our treatment allows for incoming coherent states of arbitrary intensity it is complementary to \cite{PhysRevA.82.063816}, where the scattering of low-power coherent states on a TLS is considered using a different formalism. It is also complementary to \cite{PhysRevA.85.023817}, where second-order coherence properties are examined in the case of a coherent input of arbitrary intensity.

\section*{Acknowledgments}
We thank Anton Frisk Kockum, Lars Tornberg and Per Delsing for valuable discussions. We would like to acknowledge funding from the Swedish Research Council and from the EU through the ERC and the projects PROMISCE and SCALEQIT.

\section*{References}

\bibliographystyle{unsrt}

\bibliography{tls}

\begin{thebibliography}{10}

\bibitem{clarke08}
J.~{Clarke} and F.~K. {Wilhelm}.
\newblock Superconducting quantum bits.
\newblock {\em Nature}, 453:1031--1042, 2008.

\bibitem{schoelkopf08}
R.~J. Schoelkopf and S.~M. Girvin.
\newblock Wiring up quantum systems.
\newblock {\em Nature}, 451:664--669, 2008.

\bibitem{Devoret13}
M.~H. Devoret and R.~J. Schoelkopf.
\newblock Superconducting circuits for quantum information: An outlook.
\newblock {\em Science}, 339(6124):1169--1174, 2013.

\bibitem{FornDiaz10}
P.~Forn-D\'{\i}az, J.~Lisenfeld, D.~Marcos, J.~J. Garc\'{\i}a-Ripoll,
  E.~Solano, C.~J. P.~M. Harmans, and J.~E. Mooij.
\newblock Observation of the bloch-siegert shift in a qubit-oscillator system
  in the ultrastrong coupling regime.
\newblock {\em Phys. Rev. Lett.}, 105:237001, Nov 2010.

\bibitem{niemczyk10}
T.~Niemczyk, F.~Deppe, H~Huebl, E.~P. Menzel, F~Hocke, M.~J. Schwarz, J.~J.
  Garcia-Ripoll, D~Zueco, T~H\"{u}mmer, E~Solano, A~Marx, and R~Gross.
\newblock Circuit quantum electrodynamics in the ultrastrong-coupling regime.
\newblock {\em Nature Physics}, 06, 2010.

\bibitem{knill01}
E.~{Knill}, R.~{Laflamme}, and G.~J. {Milburn}.
\newblock {A scheme for efficient quantum computation with linear optics}.
\newblock {\em Nature}, 409:46--52, 2001.

\bibitem{shen05}
Jung-Tsung Shen and Shanhui Fan.
\newblock Coherent single photon transport in a one-dimensional waveguide
  coupled with superconducting quantum bits.
\newblock {\em Phys. Rev. Lett.}, 95(21):213001, 2005.

\bibitem{shen05b}
J.~T. Shen and Shanhui Fan.
\newblock Coherent photon transport from spontaneous emission in
  one-dimensional waveguides.
\newblock {\em Opt. Lett.}, 30(15):2001--2003, 2005.

\bibitem{PhysRevA.82.063816}
Huaixiu Zheng, Daniel~J. Gauthier, and Harold~U. Baranger.
\newblock Waveguide qed: Many-body bound-state effects in coherent and
  fock-state scattering from a two-level system.
\newblock {\em Phys. Rev. A}, 82:063816, Dec 2010.

\bibitem{PhysRevA.85.043832}
Huaixiu Zheng, Daniel~J. Gauthier, and Harold~U. Baranger.
\newblock Strongly correlated photons generated by coupling a three- or
  four-level system to a waveguide.
\newblock {\em Phys. Rev. A}, 85:043832, Apr 2012.

\bibitem{PhysRevA.85.023817}
\ifmmode \mbox{\c{S}}\else \c{S}\fi{}\"ukr\"u~Ekin
  Kocaba\ifmmode~\mbox{\c{s}}\else \c{s}\fi{}, Eden Rephaeli, and Shanhui Fan.
\newblock Resonance fluorescence in a waveguide geometry.
\newblock {\em Phys. Rev. A}, 85:023817, Feb 2012.

\bibitem{peropadre13}
B~Peropadre, J~Lindkvist, I-C Hoi, C~M Wilson, J~J Garcia-Ripoll, P~Delsing,
  and G~Johansson.
\newblock Scattering of coherent states on a single artificial atom.
\newblock {\em New Journal of Physics}, 15(3):035009, 2013.

\bibitem{dibyendu10}
Dibyendu Roy.
\newblock Few-photon optical diode.
\newblock {\em Phys. Rev. B}, 81:155117, Apr 2010.

\bibitem{dibyendu11}
Dibyendu Roy.
\newblock Two-photon scattering by a driven three-level emitter in a
  one-dimensional waveguide and electromagnetically induced transparency.
\newblock {\em Phys. Rev. Lett.}, 106:053601, Feb 2011.

\bibitem{Astafiev10}
O.~Astafiev, A.~M. Zagoskin, A.~A. Abdumalikov, Yu.~A. Pashkin, T.~Yamamoto,
  K.~Inomata, Y.~Nakamura, and J.~S. Tsai.
\newblock {Resonance Fluorescence of a Single Artificial Atom}.
\newblock {\em Science}, 327(5967):840--843, 2010.

\bibitem{AstafievPRL2010}
O.~V. Astafiev, A.~A. Abdumalikov, A.~M. Zagoskin, Yu.~A. Pashkin, Y.~Nakamura,
  and J.~S. Tsai.
\newblock Ultimate on-chip quantum amplifier.
\newblock {\em Phys. Rev. Lett.}, 104:183603, 2010.

\bibitem{AbdumalikovPRL2010}
A.~A. Abdumalikov, O.~Astafiev, A.~M. Zagoskin, Yu.~A. Pashkin, Y.~Nakamura,
  and J.~S. Tsai.
\newblock Electromagnetically induced transparency on a single artificial atom.
\newblock {\em Phys. Rev. Lett.}, 104:193601, 2010.

\bibitem{AbdumalikovPRL2011}
A.~A. Abdumalikov, O.~V. Astafiev, Yu.~A. Pashkin, Y.~Nakamura, and J.~S. Tsai.
\newblock Dynamics of coherent and incoherent emission from an artificial atom
  in a 1d space.
\newblock {\em Phys. Rev. Lett.}, 107:043604, 2011.

\bibitem{hoi11}
Io-Chun Hoi, C.~M. Wilson, G\"oran Johansson, Tauno Palomaki, Borja Peropadre,
  and Per Delsing.
\newblock Demonstration of a single-photon router in the microwave regime.
\newblock {\em Phys. Rev. Lett.}, 107:073601, 2011.

\bibitem{hoi12}
Io-Chun Hoi, Tauno Palomaki, Joel Lindkvist, G\"oran Johansson, Per Delsing,
  and C.~M. Wilson.
\newblock Generation of nonclassical microwave states using an artificial atom
  in 1d open space.
\newblock {\em Phys. Rev. Lett.}, 108:263601, 2012.

\bibitem{Hoi122}
Io-Chun Hoi, Anton~F. Kockum, Tauno Palomaki, Thomas~M. Stace, Bixuan Fan, Lars
  Tornberg, Sankar~R. Sathyamoorthy, G\"oran Johansson, Per Delsing, and C.~M.
  Wilson.
\newblock Giant cross–kerr effect for propagating microwaves induced by an
  artificial atom.
\newblock {\em Phys. Rev. Lett.}, 111:053601, Aug 2013.

\bibitem{Hoi123}
Io-Chun Hoi, C~M Wilson, G{\"o}ran Johansson, Joel Lindkvist, Borja Peropadre,
  Tauno Palomaki, and Per Delsing.
\newblock Microwave quantum optics with an artificial atom in one-dimensional
  open space.
\newblock {\em New Journal of Physics}, 15(2):025011, 2013.

\bibitem{vanLoo12}
Arjan~F. van Loo, Arkady Fedorov, Kevin Lalumière, Barry~C. Sanders, Alexandre
  Blais, and Andreas Wallraff.
\newblock Photon-mediated interactions between distant artificial atoms.
\newblock {\em Science}, 342(6165):1494--1496, 2013.

\bibitem{Eichler12}
C.~Eichler, C.~Lang, J.~M. Fink, J.~Govenius, S.~Filipp, and A.~Wallraff.
\newblock Observation of entanglement between itinerant microwave photons and a
  superconducting qubit.
\newblock {\em Phys. Rev. Lett.}, 109:240501, Dec 2012.

\bibitem{menzel10}
E.~P. Menzel, F.~Deppe, M.~Mariantoni, M.~\'A. Araque~Caballero, A.~Baust,
  T.~Niemczyk, E.~Hoffmann, A.~Marx, E.~Solano, and R.~Gross.
\newblock Dual-path state reconstruction scheme for propagating quantum
  microwaves and detector noise tomography.
\newblock {\em Phys. Rev. Lett.}, 105:100401, Aug 2010.

\bibitem{blais10}
Marcus~P. da~Silva, Deniz Bozyigit, Andreas Wallraff, and Alexandre Blais.
\newblock Schemes for the observation of photon correlation functions in
  circuit qed with linear detectors.
\newblock {\em Phys. Rev. A}, 82:043804, Oct 2010.

\bibitem{romero09}
G.~Romero, J.~J. Garcia-Ripoll, and E.~Solano.
\newblock Microwave photon detector in circuit qed.
\newblock {\em Phys. Rev. Lett.}, 102:173602, Apr 2009.

\bibitem{peropadre11}
B.~Peropadre, G.~Romero, G.~Johansson, C.~M. Wilson, E.~Solano, and J.~J.
  Garcia-Ripoll.
\newblock Approaching perfect microwave photodetection in circuit qed.
\newblock {\em Phys. Rev. A}, 84:063834, Dec 2011.

\bibitem{fan13}
Bixuan Fan, Anton~F. Kockum, Joshua Combes, G\"oran Johansson, Io-chun Hoi,
  C.~M. Wilson, Per Delsing, G.~J. Milburn, and Thomas~M. Stace.
\newblock Breakdown of the cross-kerr scheme for photon counting.
\newblock {\em Phys. Rev. Lett.}, 110:053601, Jan 2013.

\bibitem{sankar13}
Sankar~R. Sathyamoorthy, L.~Tornberg, Anton~F. Kockum, Ben~Q. Baragiola, Joshua
  Combes, C.M. Wilson, Thomas~M. Stace, and G.~Johansson.
\newblock Quantum nondemolition detection of a propagating microwave photon.
\newblock {\em arXiv:1308.2208v2}.

\bibitem{houck07}
A.~A. Houck, D.~I. Schuster, J.~M. Gambetta, J.~A. Schreier, B.~R. Johnson,
  J.~M. Chow, L.~Frunzio, J.~Majer, M.~H. Devoret, S.~M. Girvin, and R.~J.
  Schoelkopf.
\newblock Generating single microwave photons in a circuit.
\newblock {\em Nature}, 449:328--331, 2007.

\bibitem{pechal13}
M~Pechal, C.~Eichler, S~Zeytinoglu, S~Berger, A~Wallraff, and S.~Filipp.
\newblock Microwave-controlled generation of shaped single photons in circuit
  quantum electrodynamics.
\newblock {\em arXiv:1308.4094v1}.

\bibitem{lang13}
C~Lang, C~Eichler, L~Steffen, J.~M. Fink, M.~J. Woolley, A~Blais, and
  A.~Wallraff.
\newblock Correlations, indistinguishability and entanglement in hong-ou-mandel
  experiments at microwave frequencies.
\newblock {\em Nat. Phys.}, 9:345--348, 2013.

\bibitem{yin13}
Yi~Yin, Yu~Chen, Daniel Sank, P.~J.~J. O'Malley, T.~C. White, R.~Barends,
  J.~Kelly, Erik Lucero, Matteo Mariantoni, A.~Megrant, C.~Neill,
  A.~Vainsencher, J.~Wenner, Alexander~N. Korotkov, A.~N. Cleland, and John~M.
  Martinis.
\newblock Catch and release of microwave photon states.
\newblock {\em Phys. Rev. Lett.}, 110:107001, Mar 2013.

\bibitem{lukin07}
Darrick~E. Chang, Anders~S. S$\o$rensen, Eugene~A. Demler, and Mikhail~A.
  Lukin.
\newblock A single-photon transistor using nanoscale surface plasmons.
\newblock {\em Nature Physics}, 03, 2007.

\bibitem{gardinerzoller}
C.~W. Gardiner and P.~Zoller.
\newblock {\em Quantum Noise}.
\newblock Springer, 1991.

\bibitem{morin13}
Olivier Morin, Jean-Daniel Bancal, Melvyn Ho, Pavel Sekatski, Virginia D'Auria,
  Nicolas Gisin, Julien Laurat, and Nicolas Sangouard.
\newblock Witnessing trustworthy single-photon entanglement with local homodyne
  measurements.
\newblock {\em Phys. Rev. Lett.}, 110:130401, Mar 2013.

\end{thebibliography}

\end{document}